\documentclass[%
 aip,
rsi,
amsmath,amssymb,
preprint,%
]{revtex4-1}

\usepackage{graphicx}
\usepackage{dcolumn}
\usepackage{bm}

\usepackage{placeins}

\usepackage[utf8]{inputenc}
\usepackage[T1]{fontenc}
\usepackage{mathptmx}
\usepackage{etoolbox}
\usepackage[caption=false]{subfig}
\usepackage{multirow}

\makeatletter
\def\@email#1#2{%
 \endgroup
 \patchcmd{\titleblock@produce}
  {\frontmatter@RRAPformat}
  {\frontmatter@RRAPformat{\produce@RRAP{*#1\href{mailto:#2}{#2}}}\frontmatter@RRAPformat}
  {}{}
}%
\makeatother

\begin{document}


\title{Flow-Mediated Regulation of Pathogen Survival in the Human Stomach}

\author{Sharun Kuhar}
    \affiliation{Department of Mechanical Engineering, Johns Hopkins University, Baltimore, Maryland 21218, USA}
\author{Jung-Hee Seo}
    \affiliation{Department of Mechanical Engineering, Johns Hopkins University, Baltimore, Maryland 21218, USA}
\author{Rajat Mittal}
    \email{mittal@jhu.edu}
    \affiliation{Department of Mechanical Engineering, Johns Hopkins University, Baltimore, Maryland 21218, USA}

\date{\today}
\begin{abstract}
\begin{center}
    \textbf{\textit{Abstract}}
\end{center}
Foodborne diseases remain a major public-health burden, and the gastric acid barrier serves as the body's primary chemical defense against ingested microbes. Yet experimentally investigating pathogen survival within this environment is highly challenging. Although recent computational stomach models have provided insights into gastric disorders, none have coupled fluid flow, acid transport, and pathogen population kinetics in a realistic stomach to assess gastric acid barrier function.
Here, we develop an imaging-based stomach model that tracks 10,000 massless particles representing pathogen colonies ingested with a liquid meal as they are advected through a dynamic, spatially heterogeneous pH field. The model incorporates acid secretion, peristaltic mixing, and gastric tone-driven emptying. Using this framework, we quantify how hypomotility and altered gastric tone influence pathogen survival.
Motility emerges as the dominant factor governing pathogen fate. The hypomotile stomach exhibits weaker mixing, retaining nearly 50\% of the initial pathogen population alive 6 minutes after ingestion, compared with less than 30\% in healthy cases. It also produces broader acid-dose distributions and more heterogeneous survival outcomes. Counterintuitively, among healthy-motility cases, increased gastric tone delivers the highest concentration of viable pathogens into the duodenum, revealing a trade-off between transport and acid-mediated inactivation.
These findings demonstrate that conventional metrics such as average pH or gastric emptying rate are insufficient for assessing gastric sterilization. Instead, the present flow–transport–kinetics framework provides new mechanistic insights into pathogen survival and gastric infection risk.
\end{abstract}

\maketitle
\section{Introduction}
The World Health Organization (WHO) estimates that 600 million cases of foodborne illnesses occur each year, out of which 420,000 result in death \cite{havelaar2015worldhealth}. In the US alone, about 48 million people suffer from foodborne illnesses every year \cite{scallan2011foodborneillness}. The low- and middle-income countries, in particular, experience an annual productivity loss of about \$95 billion \cite{jaffee2019safefood} due to these illnesses. Alongside food hygiene awareness, understanding the inadequacy of our natural defense mechanisms against contaminated foods holds immense importance.

The stomach serves as a critical first line of defense against ingested pathogens by maintaining a highly acidic environment that is hostile to microbial survival, a chemical defense commonly termed the \emph{gastric acid barrier}. However, this barrier is not always fully effective, and some pathogens can survive gastric transit, reach the intestines, and potentially cause enteric infections. Beyond food safety, pathogen survivability in the stomach is also highly relevant to the design of oral vaccines and orally administered therapeutics, where survival through the gastric environment directly impacts bioavailability and treatment efficacy.

However, predicting this survivability is a difficult endeavor. The stomach has a highly dynamic environment in which the mixing is driven by the peristaltic motion of the stomach walls, and the emptying of its contents into the intestines is driven by the contraction exerted by the proximal walls. This generates a high degree of nonuniformity in the contents of the stomach. Consequently, instead of a uniform pH environment, pathogens travel through varying pH ``microclimates'' before eventually arriving in the intestines. Whether a given initial population of pathogens survives this gastric sterilization will depend on many factors, including the rate of acid secretion, the rate of gastric emptying, food properties, and stomach wall motility.

In-vivo measurements of the survival of pathogens are procedurally and ethically challenging. In vitro setups, on the other hand, have been attempted. Barmpalia-Davis et al, 2008 \cite{barmpalia-davis2008differencessurvival} measured the population of \emph{Listeria monocytogenes} strains in a simulated dynamic gastrointestinal system. Two separate flasks, representing a stomach and an intestinal compartment, were supplied with corresponding enzymes (simulated gastric fluid and intestinal fluid with bile, respectively), and a peristaltic pump transferred contents from the stomach to the intestinal compartment, mimicking the gastric emptying rate. Additionally, an exemplary work by Koseki et al. 2010 \cite{koseki2011modelingpathogen} reported inactivation kinetics for three different pathogens at different pH values in simulated gastric fluids alongside foods. They obtained relationships between inactivation rates and pH for each pathogen. While such idealized in-vitro setups prove useful in assessing the relative survival of pathogens in uniform environments, they cannot capture the natural variations within the stomach that are generated by peristaltic contractions and regional acid secretion.

In-silico investigations of pathogen survival are scarce, even if we broaden our scope to include reduced-order models. In one study by Rahman et al. 2020 \cite{rahman2020agentbasedsimulator}, food portions and bacteria were treated as `agents' that move through the stomach at specified speeds. The agents' state was affected by the uniform but time-varying pH. They reported that 4.5 and 29\% of the ingested bacteria survived the stomach and made it to the small intestines for low and high survivability probabilities, respectively. Although the model did not account for the peristaltic motion of the walls and the fluid dynamics inside the stomach, it was a novel attempt at modeling the survivability of pathogens on their journey through the stomach. Another study by Abe et al. 2021 \cite{abe2021newdoseresponse} used a reduced-order mechanistic model to investigate the dose-response relationship between ingested bacteria and infection probability. Using a Bayesian statistical model, they estimated this probability using experimental observations of survival in a low pH environment, transition to intestines, and invasion of intestinal tissues.

Considering that there is a rapidly growing body of work demonstrating the capabilities and utilities of computational fluid dynamics in modeling stomach biomechanics \cite{palmada2023systematicreview, kuhar2024computationalmodels}, it comes as a surprise that no attempt has been made to incorporate survival kinetics into these modern approaches. Few researchers have started to investigate acid secretion and pH. Li et al. \cite{li2021mixingemptying} reported the average pH variation with different foods and different emptying rates. Even more recently, Liu et al \cite{liu2025gastricmixing} observed the importance of the rate of diffusion in radial transport of acid within a stomach with a permanently closed pylorus. However, the spatiotemporal variations in pH have not yet been coupled with bacterial population kinetics to further the discussion on what that pH variations implies for pathogen survivability.

In this study, we investigate these issues using a novel imaging-data-based, high-fidelity computational model of an anatomically realistic postprandial human stomach. The model mimics the ingestion of pathogens together with a liquid meal. Acid secreted from the gastric wall is transported throughout the stomach lumen by flow-driven convection and diffusion arising from stomach motion and pressure gradients, producing a dynamic spatiotemporal pH field. Using these simulations, we examine how variations in gastric motility and emptying rate influence pathogen transport, acid exposure, and survival.

\section{Methodology}

\subsection{Stomach Model}
The stomach geometry and motility prescription have been discussed extensively in our previous works \cite{kuhar2022effectstomach, kuhar2024silicomodelling, kuhar2025duodenogastricreflux, kuhar2025silicostudy}, but it is also briefly described here. We derive the stomach shape from the magnetic resonance (MR) imaging data from the Virtual Population Library \cite{gosselin2014developmentnew}. However, as it depicts a low-food volume stomach, it was modified to resemble a postprandial stomach based on publicly available MR images from sources such as the study by Lu et al \cite{lu2022automaticassessment} and the website of Motilent (London, UK). The geometry consists of the stomach lumen as well as a portion of the duodenum (the initial segment of the small intestine) that receives the contents emptied by the stomach.  A triangulated mesh of 31,648 elements with edge size $\sim1$ mm is used to discretize the geometry. The parameters for the peristaltic motion of the stomach walls are obtained from existing imaging studies that report the amplitude, speed, and frequency of contraction waves along with the spatial variation of these parameters as they propagate from the proximal end (near the esophagus) towards the distal end (towards the duodenum). We use 40 mm wavelength peristaltic waves that originate in the proximal stomach every 20 s, growing from 0\% occlusion in the proximal stomach to 40\% in the antrum while moving at a speed of 2.3 mm/s \cite{ferrua2010modelingfluid, kuhar2022effectstomach}. Consequently, at any instant, there are three simultaneous peristaltic contractions in tandem moving towards the pylorus - a sphincter that connects the stomach to the duodenum. As a contraction approaches the pylorus, the orifice opens up to a diameter of 2 mm \cite{kong2008disintegrationsolid}. When the contraction finally reaches the pylorus, it terminates with a large amplitude collapse of the stomach walls, known as the terminal antral contraction \cite{schulze2006imagingmodelling,camilleri2015gastricmotility}. The pyloric orifice closes before the collapse is complete, resulting in the trapped contents being regurgitated back into the stomach, achieving enhanced mixing and trituration. The timing of the pyloric opening and closing was informed using the approach described by Ishida et al \cite{ishida2019quantificationgastric}.

In addition to the kinematics of the stomach, our model also accounts for gastric tone, i.e., the force exerted by the upper stomach walls on the contents, which is known to be the primary driver of gastric emptying \cite{camilleri2015gastricmotility}. The push exerted by the proximal stomach is modeled through a boundary condition\cite{kuhar2025duodenogastricreflux} that mimics this behavior as described ahead.

\subsection{Fluid Flow Model}
The complex shape and motion of the stomach geometry are simulated by immersing it into a Cartesian mesh.  The computational domain is of size $15\times10\times13$ cm$^3$, discretized by cubical elements of edge size 0.5 mm, chosen based on a grid independence study in a prior study \cite{kuhar2022effectstomach}. The stomach boundary is resolved using the immersed boundary method in which the equations are solved only in the fluid region (i.e., inside the stomach and duodenum for the current geometry) and the boundary condition is imposed by specifying the velocity in the neighboring solid cells \cite{mittal2008versatilesharp}. 

The contents are assumed to be aqueous ($\mu=1$ mPa s, $\rho=1000$ kg/m$^3$), whose motion is governed by the incompressible Newtonian fluid governed by the Navier-Stokes equations:
\begin{align}\label{eq:ns}
    \bm{\nabla} \cdot \mathbf{u} &= 0,\\
    \rho\left( \frac{\partial \mathbf{u}}{\partial t} + \left(\mathbf{u} \cdot \bm{\nabla}\right) \mathbf{u}\right) 
        &= -\bm{\nabla} p + \mu {\nabla}^2 \mathbf{u}.
\end{align}
The solver employs the fractional-step method to advance these equations in time \cite{chorin1969convergencediscrete}. A second-order central differencing scheme is used for the advection and diffusion terms, while the Crank-Nicolson scheme is used for time integration, with the non-linear advection term being updated iteratively until convergence. A stabilized bi-conjugate gradient method is used to solve the pressure Poisson equation \cite{zhu2018computationalmodelling}.

A no-slip boundary condition ($\mathbf{u} = \bm{0}$) is applied along the walls of the stomach and duodenum. An outflow boundary condition ($\partial \mathbf{u}/\partial n = \bm{0}$) is applied at the duodenum exit. At the fundus opening, a boundary condition inspired by the fundic accommodation is implemented (see \cite{kuhar2025duodenogastricreflux}): $d \bar{u}/dt = (P_o-\bar{p})/M$, where $\bar{u}$ is the specified flow velocity normal to the boundary surface and $\bar{p}$ is the average fluid pressure at the boundary. The boundary condition also contains two parameters: $P_o$ represents the pressure exerted by the fundus/proximal-stomach on the contents, while $M$ is a stabilizing factor that represents the inertia of the system. If the fluid pressure is lower than the specified fundic pressure, fluid flows in from the remaining stomach region that lies outside the computational domain, and in case of the opposite, it flows back. For the healthy case, we choose $P_o$ to be the value which results in the gastric emptying rate observed in low-calorie liquid meals (4.1 mL/min \cite{brener1983regulationgastric, marciani2001effectmeal}). To model scenarios of lower or higher gastric tone, we modify $P_o$ to half or double the healthy value. The inertia parameter $M$ is kept constant at 1000 throughout the study, as it helped stabilize the simulation for all cases, while the results remained insensitive towards even a 5x change in the value of $M$.

\subsection{pH Model}
Mathematically, pH is defined as:
\begin{equation}
    \textrm{pH} = -\log_{10}[\textrm{H}^+],
\end{equation}
where $[\textrm{H}^+]$ is the concentration of hydrogen ions. The model treats hydrogen ions as a passive scalar dissolved in the fluid medium. The advection-diffusion equation is solved to track the transport of ions across the stomach:
\begin{equation}\label{eq:ad}
    \frac{\partial c_{\text{H}^+}}{\partial t}+ \left(\mathbf{u} \cdot \bm{\nabla}\right) c_{\text{H}^++} = D\vec{\nabla}^2 c_{\text{H}^+},
\end{equation}
where $c_{\text{H}^+}$ is the concentration of ions ($c_{\text{H}^+} = [\textrm{H}^+]$), $\mathbf{u}$ is the fluid velocity, and $D = 3\times10^{-9}$ m$^2$/s is the coefficient of diffusion of the ions \cite{lobo1979diffusioncoefficients, trusov2016multiphaseflow}. To solve this partial differential equation, a second-order central-differencing scheme is used to discretize the diffusion terms, whereas a hybrid central and upwind scheme is used for the advection term. The stomach walls are specified with a flux boundary condition ($\text{D}\, \partial c_{H^+}/\partial n = \dot{q}$) such that the rate of total flux ($\dot{Q} = \int \dot{q} dS$) matches the experimentally observed acid secretion rate in humans. We used a total acid flux of $\dot{Q}=$ 10 mEq H$^+$/h \cite{heuman1997gastroenterology, bornhorst2014gastricdigestion} uniformly secreted only from the proximal region \cite{bornhorst2014gastricdigestion} of the stomach with an area of 182 cm$^2$. The remaining surface area of the stomach has no acid secretion and was specified with $\dot{q} = 0$.

\subsection{Pathogen Model}
Due to their microscopic size compared to the length scales in the flow problem, the pathogens can safely be treated as massless point particles. Furthermore, we treat each particle not as an individual pathogen, but as a collection of colony-forming units of pathogens. This allows us to statistically monitor the pathogen population without having to decide the fate of each pathogen individually. The trajectory of each colony is then decided by the local fluid velocity at the location of the particles:
\begin{equation}\label{eq:path}
    \frac{\text{d}\bm{x}_p}{\text{d}t} = \mathbf{u}(\bm{x_p}),
\end{equation}
where $\bm{x}_p (t)$ is the location of a particle and $\mathbf{u}$ is the fluid velocity field. For each particle, the above ordinary differential equation is solved explicitly in time using a fourth-order Runge-Kutta method. The pH exposure history of each particle is also stored at each time step to inform the population model: $\text{pH}_p(t) = \text{pH}(\bm{x}_p(t))$.

The experiments with pathogens in simulated gastric fluid and the population models presented by Koseki et al \cite{koseki2011modelingpathogen} are used to calculate the population over time of each colony. The population, $N_p(t)$, of a colony is described by the modified logistic model:
\begin{equation}\label{eq:popn}
    \frac{\text{d}N_p}{\text{d}t} = -k_{\text{max}}(\text{pH}_p)\left[ 1 - \left(\frac{N_{\text{min}}}{N_p}\right)^m \right]N_p,
\end{equation}
where $k_{\text{max}}$ is the maximum specific inactivation rate and is dependent on the local pH, $N_{\text{min}}$ represents the minimum cell population at the end of the experiment, and $m$ characterizes the curve shape. This population model captures an exponential death phase followed by a tail \cite{koseki2011modelingpathogen}. Among the various pathogens for which experimental measurements are available, we chose \emph{S.} Tymphimurium as it shows increased gastric sensitivity \cite{akritidou2022effectgastric} while also being able to survive for prolonged duration in low pH environments \cite{jonge2003adaptiveresponses}. In accordance with the experimental data \cite{koseki2011modelingpathogen}, $N_{\text{min}}$ is set to 1 CFU/mL, $m$ is fixed at 0.1, and the correlation between $k_{\text{max}}$ and pH is given by:
\begin{equation}
    \sqrt{k_{\text{max}}} = -1.15\cdot(\text{pH} - 2.85).
\end{equation}
At every time step of the flow solver, for each particle, the ordinary differential equation, Equation \eqref{eq:popn}, is marched explicitly in time using a forward Euler scheme based on the local pH. The population decreases in accordance with the local pH as the particle (representing a colony) propagates through the stomach until it crosses the pylorus, when the remaining population is counted as survivors that reached the duodenum.

\subsection{Simulation Procedure}
\begin{figure}
    \centering
    \includegraphics{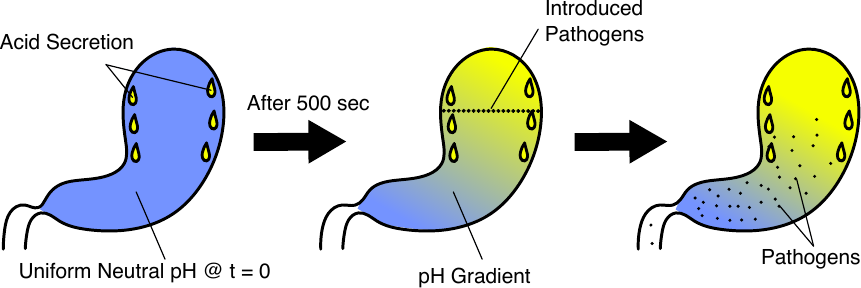}
    \caption{Schematic describing the simulation procedure. First, a typical pH gradient is established by mixing the acid secretion from the proximal walls. Then the pathogens are introduced in a layer in the upper stomach, which subsequently gets transported through the varying pH environment.}
    \label{fig:methodschematic}
\end{figure}
Introducing pathogens in a uniformly neutral or uniformly acidic stomach environment would fail to capture the typical pH gradients observed inside the stomach \cite{simonian2005regionalpostprandial}. To ensure a more typical pH environment for pathogens, the simulations were carried out in two phases. In the first phase, we start with a neutral pH ($\textrm{pH} = 7$) throughout the stomach and solve the flow (Equation \ref{eq:ns}) and transport (Equation \ref{eq:ad}) equations for 500 s of physical time. During this time, 25 peristalsis waves finished their journey across the stomach and developed a typical spatially varying acidic environment expected in the early postprandial phase of gastric digestion.

In the second phase, we resume the simulation and introduce pathogens in a uniformly distributed layer in the upper stomach to mimic a scenario right after someone ingests harmful pathogens along with the meal. Although a uniform distribution in a single layer is an idealized scenario, the motive behind it was to eliminate confounding effects associated with the initial location of arrival of pathogens into the stomach, which is not the focus of the current study. Alongside the flow and transport equations, the particle motion (Equation \eqref{eq:path}) and population (Equation \eqref{eq:popn}) equations are also solved in this second phase, as we must evaluate the trajectory of particles as they pass through a spatially varying pH environment.

We simulate a total of four cases in this study. The control case has a healthy motility amplitude of 40\% occlusion \cite{ferrua2010modelingfluid} and a healthy gastric tone corresponding to a fundic pressure of 0.1 mm Hg, which is based on the observed emptying rate of 4.1 ml/min for low-calorie liquid meals \cite{brener1983regulationgastric, marciani2001effectmeal}. We also study a hypomotility case with 50\% lower peristalsis wave amplitude (i.e., an occlusion of 20\%), which represents a scenario where the motility has weakened due to idiopathic gastroparesis \cite{grover2019gastroparesisturning}, diabetes-induced gastroparesis \cite{ajaj2004realtime, urbain1993characterizationgastric}, or other conditions such as Parkinson's disease \cite{hardoff2001gastricemptying}. The other two cases incorporate variations in gastric tone, which is known to be the primary driver of gastric emptying. It can be increased or decreased by diet \cite{livzan2023diagnosticprinciples}, medications \cite{lee2010metoclopramidetreatment}, hormones \cite{cuomo2006influencemotilin} and glucagon-like peptide-1 (GLP-1) drugs \cite{hellstrom2001interactionsgastric}. To account for both an increase as well as a decrease in gastric tone, we simulated two cases with 50\% higher and lower fundic pressure values ($P_o$) with respect to the control.

\section{Results and Discussion}

\subsection{pH Distribution}
\begin{figure}
    \centering
    \includegraphics[width=0.7\linewidth]{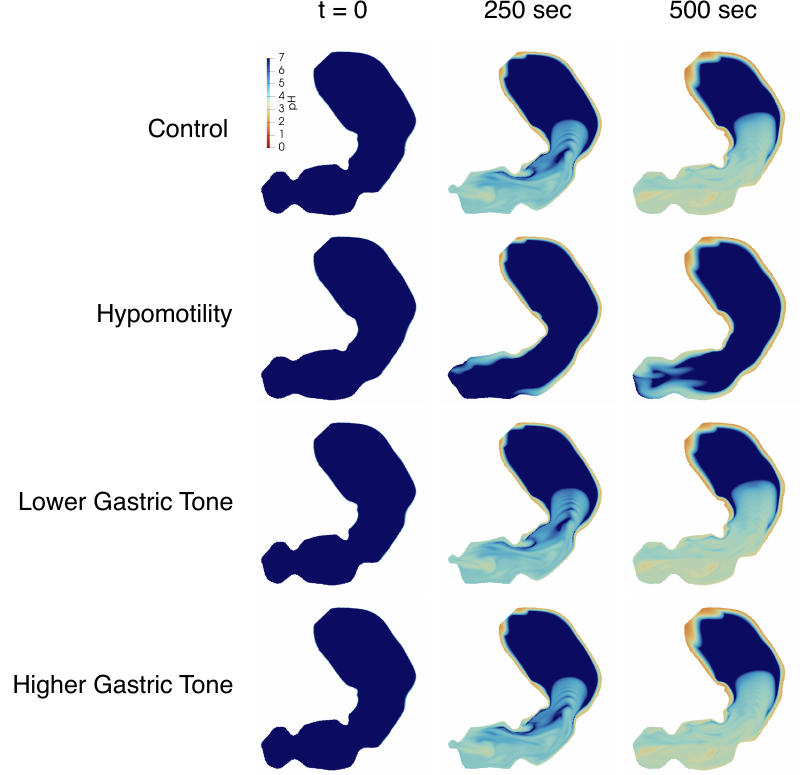}
    \caption{The complex pH distribution shown on a slice across the stomach for different cases.}
    \label{fig:pH_on_slice}
\end{figure}

Before the pathogens are introduced into the proximal stomach, each of the four cases evolves from a uniform neutral pH environment to form a typical pH distribution as shown by the pH contours on a slice through the lumen in Figure \ref{fig:pH_on_slice}. With the acid secretion rate kept constant, the distribution is governed by the motility and gastric tone. The hypomotility case exhibits the most limited distribution of the acidic regions, while the other three cases qualitatively look similar in their ability to widely distribute the acid secreted from the walls through the lumen of the stomach. 

\begin{figure}
    \centering
    \includegraphics[width=0.5\textwidth]{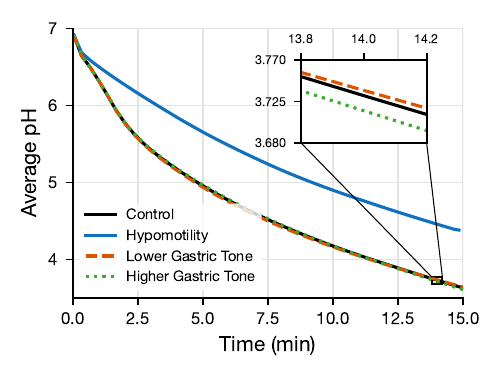}
    \caption{The variation of average pH inside the stomach.}
    \label{fig:avgpH}
\end{figure}
In Figure \ref{fig:avgpH}, the variation of the volume-averaged pH over time is compared for all cases. As observed in the qualitative distribution, hypomotility consistently remains the least acidic in terms of the average pH, while the other three cases stay close together. The differences between the three similar cases are revealed by the inset: higher gastric tone has a lower average pH compared to the control case, which is less than that of the lower gastric tone case.

These results provide two key insights. First, motility has a dominant role in mixing the contents of the stomach and the gastric secretions. Any variations in the gastric tone, which is the primary determinant of the emptying rate, are negligible compared to the effects of changes in mixing due to motility. In other words, healthy stomach motility is much more important in achieving a more uniform and acidic environment than gastric tone. Secondly, contributions by a higher gastric tone, albeit much smaller, further lower the pH because the overall increase in emptying rate helps in faster transport of proximal contents (closer to the esophagus) into the distal stomach (close to the pylorus and intestines).

\begin{figure}
    \centering
    \includegraphics[width=0.5\textwidth]{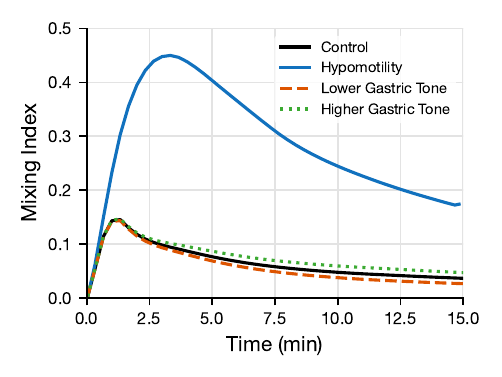}
    \caption{Mixing Index of the stomach for different cases. A value of 0 indicates a perfect mixture, and 1 indicates completely segregated contents.}
    \label{fig:mixing}
\end{figure}

A more traditionally used quantity to capture the stomach's efficacy of mixing acid with the rest of the contents is the mixing index \cite{bornhorst2014gastricdigestion}. Here, we calculate the mixing index based on the intensity of segregation definition suggested by Luo et al \cite{luo2006studymixing,fu2017mixingindexes},
\begin{equation}
\text{mixing index}=\frac{{\frac{1}{n-1}\sum\left(c_{\text{H}^+}-\bar{c}_{\text{H}^+}\right)^2}}{\bar{c}_{\text{H}^+}(1-\bar{c}_{\text{H}^+})}.
\end{equation}
The mixing index over time is shown in Figure \ref{fig:mixing}, with a value of 0 denoting a perfect mixture. At $t=0$, there is no acid inside the stomach in all cases and, correspondingly, the mixing index is at 0. As acid is secreted from the proximal walls, the non-uniformity in the concentration field is noted by a rising mixing index followed by a decline, which corresponds to the mixing of secreted acid with the contents over time. The hypomotility case remains the most poorly mixed case, while the remaining three cases exhibit mixing indices that appear in order of their emptying rates. The lower gastric tone allows the contents to mix better than the control case, which in turn mixes better than the higher gastric tone case.

\subsection{Pathogen Distribution}
\begin{figure}
    \centering
    \includegraphics{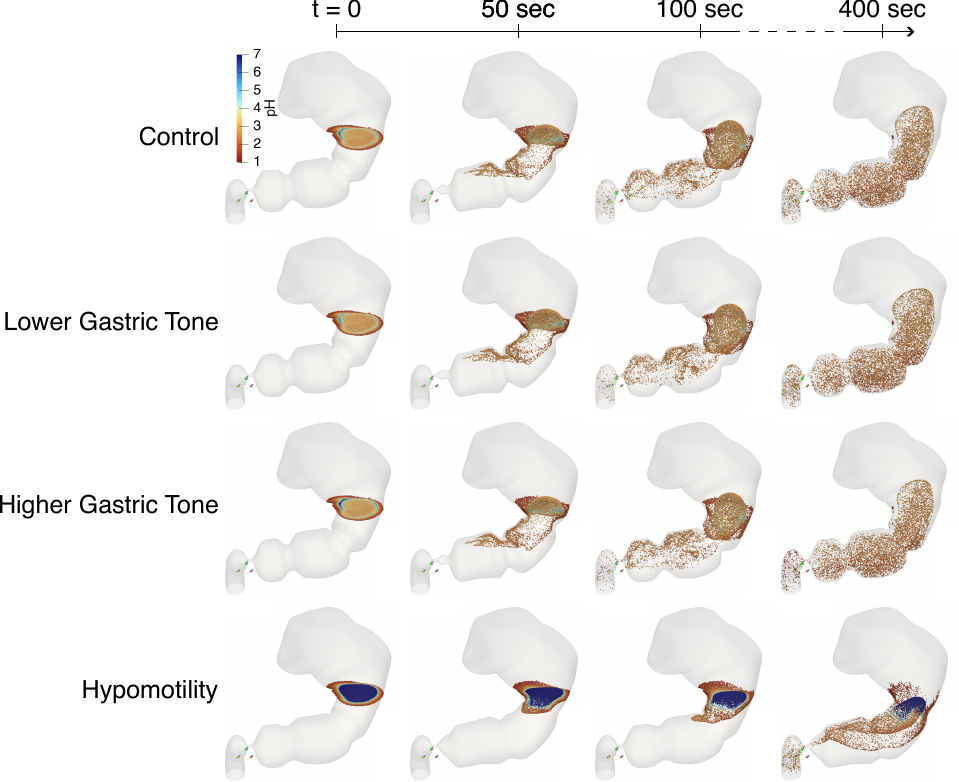}
    \caption{Distribution of pathogens across the geometry over time for different cases.}
    \label{fig:parts}
\end{figure}

After a typical pH distribution is established, the pathogens are introduced by distributing them uniformly over a plane into the proximal stomach. The local flow velocity moves the pathogens across the stomach lumen, exposing them to a varying pH atmosphere. Figure \ref{fig:parts} shows the distribution of pathogens over time, with the color denoting the pH at the pathogen location.

As seen by the persistent high pH particles (blue-colored in Figure \ref{fig:parts}), the poor transport in the case of the hypomotile stomach translates to a majority of pathogens continuing to circulate in a high pH environment, even more than 6 minutes after the introduction of pathogens. The low-amplitude waves fail to move the pathogen particles that lie at the center of the lumen. On the other hand, the qualitative differences between control and lower/higher gastric tone cases are subtle. For example, local high pH regions are more pronounced in the high motility case due to faster emptying of contents, giving them less time to mix with the secreted acid. However, a more quantitative representation is needed to further analyze the differences between these cases.

\subsection{Acid Exposure of the Total Population}
\begin{figure}
    \centering
    \includegraphics[width=0.5\textwidth]{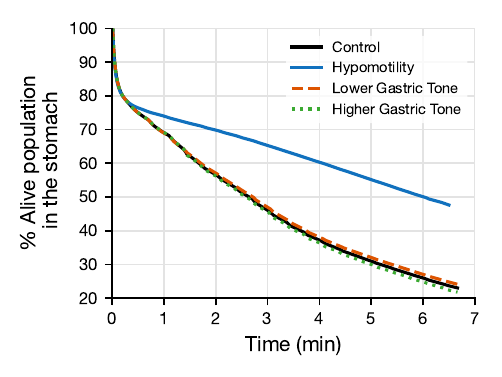}
    \caption{Percent of alive pathogens inside the stomach over time.}
    \label{fig:aliveinstom}
\end{figure}

Before discussing pathogens that reach the duodenum, we must first look at the population evolution inside the stomach because only focusing on the duodenum population would overlook those that could not empty the stomach within the duration of the simulation. A significant portion of the microbial load remains within the stomach and would empty over the subsequent hours. To assess their survivability trends, we show the total number of pathogens within the stomach that are alive in Figure \ref{fig:aliveinstom}. The figure shows that healthy stomach motility, irrespective of gastric tone, has managed to kill over 70\% of the initial pathogens, while a hypomotile stomach performs the worst with the maximum number of pathogens alive.

\begin{figure}
    \centering
    \includegraphics[width=0.5\textwidth]{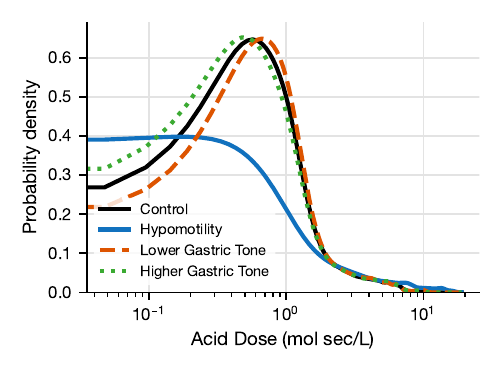}
    \caption{Probability density function of the acid dose of the population.}
    \label{fig:pdfn}
\end{figure}

We also calculate the acid dose that all particles (the ones that could empty as well as the ones that could not) are exposed to by:
\begin{equation}
    \text{Acid Dose} = {\int_0^{t_{empt}} c_{\text{H}^+}(\bm{x}_p)\, d\tau }
\end{equation}
where $t_{empt}$ is the time at which the particle is emptied from the stomach. If by the end of the simulation a particle stays within the stomach, then $t_{empt}$ is equal to the end time of the simulation. The probability density function of the acid dose that the 10,000 particles were exposed to is shown in Figure \ref{fig:pdfn}. This figure plots the fraction of the population exposed to different acid doses for all four cases. The hypomotility case is revealed to be the most dangerous, with a high fraction of the pathogens having a low acid dose, increasing the likelihood of their survival. The effect of gastric tone is also revealed with the curve shifting leftwards with increasing gastric tone - implying a higher fraction of pathogens being exposed to low acid dose with increasing emptying rate.

\subsection{Pathogen Population Arriving in the Duodenum}
\begin{figure}
    \centering
    \subfloat[]{\includegraphics[width=0.5\textwidth]{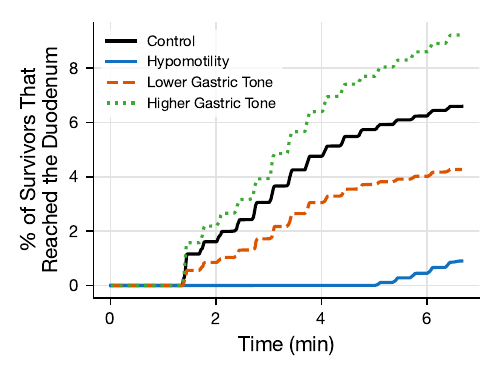}}\hfill
    \subfloat[]{\includegraphics[width=0.5\textwidth]{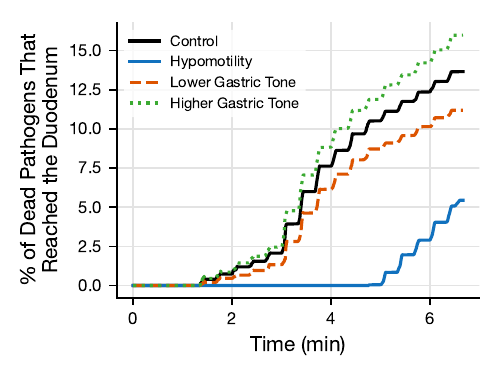}}
    \caption{Figures showing the \% of (a) survivors, and (b) dead pathogens that reached the duodenum over time. The percentages are calculated with respect to the initial total population of the pathogens introduced into the stomach.}
    \label{fig:duodenum_population}
\end{figure}
\begin{figure}
    \centering
    \includegraphics[width=0.5\textwidth]{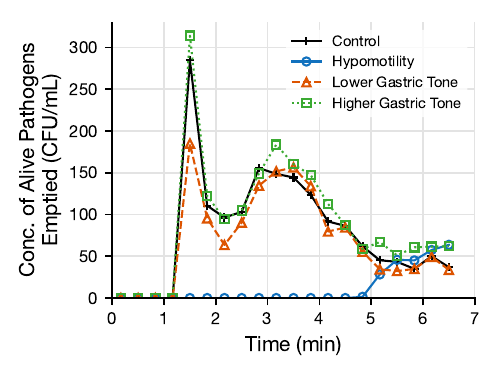}
    \caption{Concentration of alive pathogens emptied with the bolus over time.}
    \label{fig:survconc}
\end{figure}

Finally, the percentage of the pathogens that survived their journey to the duodenum with respect to the total initial population is shown in Figure \ref{fig:duodenum_population}(a). These would be the pathogens that have the potential to cause enteric infections. The adjacent Figure \ref{fig:duodenum_population}(b) denotes their counterpart: the percentage that are emptied but could not survive the journey. The staircase pattern in both figures is a consequence of stomach contents emptying in bursts as the pylorus opens up every 20 seconds in synchrony with the incoming contractions.  Additionally, in Figure \ref{fig:survconc} we show the concentration of alive microbes delivered to the duodenum to eliminate the confounding effects associated with the distinct emptying rate of different cases. It is revealed that, in the beginning, a higher gastric tone proves to be the most harmful as it delivers the highest concentration of live microbes in the early stages, while all cases behave similarly towards the end of the simulation.

To quantify the bactericidal effectiveness of each case, we compute the fraction of pathogens that are inactivated prior to emptying into the duodenum, normalized by the total number of pathogens emptied. This comparison, summarized in Table \ref{tab:erwithwithout}, shows that the hypomotile case exhibits the highest antimicrobial potency, despite its weaker mixing and slower transport. The increased residence time within the stomach allows a larger fraction of emptied pathogens to be exposed to lethal acid doses before reaching the duodenum. In contrast, the high-gastric-tone case exhibits the lowest bactericidal effectiveness because accelerated emptying transports pathogens into the duodenum before sufficient acid exposure can occur.
\begin{table}
\centering
\caption{A comparison of the efficacy of the stomach in killing ingested pathogens for different cases. Antimicrobial Potency is defined as $ = 1 - $ Alive Pathogens/Total Pathogens reaching the duodenum.}
\begin{tabular}{cc}
\textbf{Case} & \textbf{Antimicrobial Potency}\\
\hline
Control & 0.67\\
Hypomotility & 0.86\\
Lower Gastric Tone & 0.72\\
Higher Gastric Tone & 0.63\\
\hline
\end{tabular}
\label{tab:erwithwithout}
\end{table}

\section{Conclusions}

In this study, we developed a high-fidelity, anatomically realistic computational framework to investigate gastric pathogen survivability by coupling fluid mechanics, acid transport, and population kinetics. By integrating peristaltic motility, gastric emptying, and spatially heterogeneous acid secretion, the model captures the complex spatiotemporal pH environment encountered by an ingested pathogen; this is an aspect that has not been addressed in prior in vitro or reduced-order approaches.

The results highlight the dominant role of stomach motility in regulating both acid mixing and pathogen exposure. Hypomotility, by virtue of lower amplitude of the peristalsis waves, significantly impairs mixing, leading to persistently higher pH (lower acidity) regions and slower transport of pathogens towards the pylorus. Interestingly, while this condition increases the residence time of the pathogens inside the stomach, there is higher variability in the distribution of acid dose that the pathogens are exposed to - a direct corollary of the poor mixing. In other words, despite staying in the stomach longer, some pathogens may get exposed to enough acid to ensure they don't survive, but many others may persist for long durations in higher pH regions.

In contrast, with a healthy stomach motility, variations in gastric tone have a characteristically distinct influence on pathogen survivability. Higher gastric tone primarily accelerates gastric emptying, increasing the microbial load reaching the duodenum. Since gastric tone had little effect on mixing, the faster rate of emptying implied that a higher proportion of pathogens would spend less time in a well-mixed stomach. A higher gastric tone consistently delivered the highest concentration of viable pathogens into the duodenum over the course of the simulation.

These findings underscore a subtle but important trade-off between transport and inactivation of pathogens in the stomach: faster emptying promotes pathogen passage, while slower transport enhances \emph{in-situ} inactivation. This effect exists in juxtaposition to the role of motility in mixing the secreted acid with the contents, without which many pathogens can continue to survive in pockets of high pH for long durations. The results also suggest that conventional metrics such as average pH or emptying rate alone are insufficient to assess gastric acid barrier function; instead, a coupled analysis of flow, transport, and reaction kinetics is essential.

Overall, this work provides a mechanistic basis for understanding variability in pathogen survival due to physiological differences in gastric function. The framework can be extended to assess food-dependent effects, non-Newtonian meal rheology, and targeted delivery of oral therapeutics. More broadly, it establishes a foundation for predictive, physics-based modeling of gastrointestinal processes with direct implications for food safety, infection risk assessment, and drug delivery design. 

\section*{Acknowledgment}
We would like to acknowledge research funding from the National Science Foundation (NSF) via Award No. CBET 2019405.

\bibliography{bibfile}

\end{document}